\documentstyle[12pt]{article}
\def\beq{\begin{equation}}
\def\eeq{\end{equation}}
\def\bea{\begin{eqnarray}}
\def\eea{\end{eqnarray}}

\begin{document}
\baselineskip .80cm
\begin{titlepage}
\begin{center}
\hfill SNUTP 93-56\\
\null\hfill \\
\vskip 0.9cm
{\large \bf
Is Heavy Quark Axion Necessarily Hadronic Axion?}
\vskip 0.9cm
{Sanghyeon Chang \ and \ Jihn E. Kim}
\\
{\it Department of Physics and Center for Theoretical Physics}\\[-0.3cm]
{\it Seoul National University, Seoul 151-742, Korea}

\vskip 1.0cm

{\bf Abstract}
\end{center}
\begin{quotation}

We show that heavy quark axion is not necessarily a hadronic axion,
which manifests in the quark and lepton seesaw
mechanism. We introduce a heavy $SU(2)$
singlet fermion for  each known fermion in order to unify the
axion scale and the seesaw scale. The light quarks and leptons
gain their masses by  the seesaw mechanism.
Even though our axion model gives a kind of heavy quark axion,
the axion has tree level lepton--axion coupling suppressed
by $F_a$, contrary to a widely known belief that heavy quark
axions are hadronic.
\end{quotation}
\end{titlepage}
\setcounter{page}{2}

$\theta_{QCD}$ is one of twenty  parameters of the standard
model which is expected to be explicable in a satisfactory theory,
which is the well--known strong CP problem.
Axion is the most attractive solution for the strong CP problem
\cite{pq,kim}.
At present there exist two viable axion models.
One is the heavy quark axion \cite{axion1} and the other is the
Dine--Fischler--Srednicki--Zhiitnisky (DFSZ) axion \cite{axion2}.
Usually heavy quark axion is referred to as hadronic axion because
the heavy quark axion has no tree level coupling with the leptons. For
the DFSZ model there exists the lepton--axion
couplings of order $m_l/F_a$.
Except for this lepton couplings, the heavy quark axion and the DFSZ
axion models have hadronic couplings of the same order.  Thus most
phenomenological and astrophysical
consequences of these two models are similar\footnote{Actually, there has
been several arguments about cosmological difference between the DFSZ and
the hadronic axion\cite{kim,kt}.}.   The invisible axion
lifetime is much larger than the age of the universe, and hence
the classical axion field can contribute to the mass density
of the universe significantly.  This consideration gives a bound on
the axion decay constant, $F_a\leq 10^{12}$ GeV; $F_a\sim 10^{12}$
GeV can supply cold axions as the needed cold dark matter.

In the standard model, neutrinos are massless.  However, the solar
neutrino problem invites a speculation for tiny neutrino masses,
$\Delta m_\nu\sim$ O(eV).  An attractive suggestion for this
tiny mass is the so--called seesaw mechanism \cite{seesaw}, by
introducing $SU(2)\times U(1)$ singlet(s) at an intermediate
scale $10^{10-15}$ GeV.

The two scales encountered above falls in the common region;
$10^{10}-10^{13}$ GeV\footnote{We extended the upper bound of
axion scale a bit, because
there exists a possibility of raising it as noted in the saxino
cosmology\cite{saxino}.}  This common scale appears in supergravity models
also \cite{mu}.  Therefore, it is quite intriguing to speculate
that the axion scale and the seesaw scale arise from the same
origin.  In fact, it is easy to relate these two scales
in grand unified models \cite{gut}.  If the symmetry of the grand
unification group $G$ is $G\times U(1)_{PQ}$, the Peccei--Quinn
symmetry breaking scale $F_a$ is the axion scale.  If $G$ breaks
down to the standard model gauge group at the grand unification
scale $M_X$, there may exist some $SU(3)\times SU(2)\times U(1)$
singlets which remain massless at $M_X$, protected by $U(1)_{PQ}$.
These singlets can obtain masses when the Peccei--Quinn symmetry is
broken at the axion scale; thus the axion scale becomes the
seesaw scale, and the Chikashige--Mohapatra--Peccei majoron\cite{pec}
is the same as the invisible axion.

Existing models which try to unify
these two scales are based on the DFSZ axion model;
two Higgs doublet and one singlet scalar \cite{gut,pec} are introduced
at the level of the standard model.

In this paper, we study the unification of these two mass
scales in the framework of the heavy quark axion model.  Namely,
instead of introducing
a new Higgs doublet, we introduce a heavy singlet quark which
couples to the same singlet scalar as
singlet right-handed neutrino in the seesaw model. Thus there
exists only one Higgs doublet at low energy.
We can set a universal model for solving
the strong CP problem and the solar neutrino puzzle based on  the heavy
quark axion model.
To make a consistent model, we need to introduce a heavy singlet
partner to each lepton.
In grand unified theory (GUT), we encounter frequently heavy
$SU(2)\times U(1)$ singlet fermions below the GUT scale. [In superstring
standard models, these singlet leptons and quarks are almost
inevitable \cite{hbk}.]   If heavy leptons arise as
singlets below the GUT scale, there may exits heavy
$SU(2)\times U(1)$ singlet quarks
also \cite{hbk}.    So it is not unreasonable to
 assume that all light fermions have
their heavy singlet partners and get their masses by
the seesaw mechanism only.
At low energy scale, phenomenology of the quark sector seesaw model is the
same as the standard model except for the axionic interactions.
This axion is an invisible heavy quark axion but {\it not} hadronic axion.
And the quark axion interaction in this model is different from
that of the hadronic axion.
This illustrates that the heavy quark axion model
is not necessarily a hadronic axion, which makes it more difficult
to rule out heavy quark axions by finding out axion--electron coupling.
On the other hand, it is a good news for experiments; more invisible
axion models have electron couplings, and probing the electron coupling
in the axion search experiments applies to wider classes of models.

In addition to the familiar quarks and leptons in the standard model,
we postulate
new $SU(2)$ singlet fermions and one complex scalar $\sigma$ \cite{kim}.
The fermion contents of the model is,
\renewcommand{\theequation}{1.\alph{equation}}
\bea
\mbox{in the quark sector }&:&q^i_L,u^i_R,d^i_R, U^i_L,U^i_R,D^i_L,D^i_R,
\nonumber\\
\mbox{in the lepton sector}&:&l^i_L,e^i_R,N^i_R,E^i_L,E^i_R,
\eea
and an extra heavy quark to realize the heavy quark axion,
\beq
Q_L, Q_R,
\eeq
where $i=1,2,3$ is a family index. $q_L^i$ and $l_L^i$ are $SU(2)$
doublets, and the other fields are $SU(2)$ singlets.
Extra particles added to the standard
model are denoted as capital letters. Their electromagnetic
charges are the same as those
of the corresponding light particles (lower case symbols)
but the electromagnetic
charge of $Q$ is undetermined.
We assign PQ charges to fermions as,
\renewcommand{\theequation}{\arabic{equation}}
\setcounter{equation}{1}
\bea
1&{\rm for }& Q_R, U_R, D_R, E_R, N_R, q_L, l_L,\nonumber\\
-1&{\rm for }& Q_L, U_L, D_L, E_L, u_R, d_R, e_R,\nonumber\\
0&{\rm for}& H,\nonumber\\
-2&{\rm for }& \sigma .
\eea
Note that the Higgs doublet carries vanishing PQ charge.

Then the Lagrangian can be written as
\bea
{\cal L}&=&b_\sigma \sigma \bar{Q}_L Q_R + b_U\sigma
\bar{U}_LU_R+b_D\sigma\bar{D}_LD_R+b_E\sigma \bar{E}_LE_R+
\frac{1}{2}b_N\sigma\bar{N}^c_R N_R\nonumber\\
&&+h_U\bar{q}_L\tilde{H}U_R +h_D\bar{q}_LHD_R
+h_N\bar{l}_L\tilde{H}N_R +h_E\bar{l}_LHE_R
\nonumber\\
&&+\alpha_U \bar{U}_Lu_R+\alpha_D\bar{D}_Ld_R
+\alpha_E\bar{E}_Le_R+{\rm h. c.}\nonumber\\
&&-V(\sigma,H)+{\cal L}_{kinetic}+{\cal L}_{gauge}
\label{1},\eea
where $H$ is a Higgs doublet scalar, ${\tilde H}=i\sigma_2 H^*$,
$b_N$ is a Hermitian matrix while the other coupling matrices
are complex,
and
\beq
V(\sigma,H)=\mu^2_HH^\dagger H+\mu^2_\sigma\sigma^*\sigma+
\lambda_H(H^\dagger H)^2+\lambda_\sigma(\sigma^*\sigma)^2+
\lambda_{\sigma H}H^\dagger H\sigma^*\sigma
.\eeq
After the spontaneous symmetry breaking, one has
\beq
\sigma=\frac{\tilde v+\tilde\rho}{\sqrt{2}}e^{ia/\tilde v},\ \ \
H=\frac{1}{\sqrt{2}}\left(\begin{array}{c}0\\v+\rho\end{array}\right)
e^{i\phi/v}.
\eeq
Here $\tilde v$ is the Peccei--Quinn symmetry breaking scale
and $v$ is the weak scale.
The tree level mass matrix of fermions except for neutrinos written
in the  $(f^1,f^2,f^3,F^1,F^2,F^3)_L$ $\otimes
(f^1,f^2,f^3,F^1,F^2,F^3)_R$ space is\footnote{There are three such
matrices, for $Q_{em}=0,2/3,-1/3$, respectively.}

\beq
M_f=\left(\begin{array}{cc}0&h_Fv\\ \alpha_F&b_F\tilde v\end{array}\right),
\eeq
where $b,\ h,\ \alpha$ are $3\times3$ matrices and $f$
and $F$ represent approximately light and
heavy fermions, respectively.
To simplify further analysis, assume $b_F$ and $\alpha_F$ are diagonal so
that $h_F$ can be diagonalized by biunitary transformation.  Namely, we
are neglecting the mixing angles.  If one of the three matrices,
$\alpha_F, b_F$ and $h_F$, is diagonal, the mixing can be introduced
in general.
So we can separate  single family mass matrix
\beq
M^i_f=\left(\begin{array}{cc}0&h^i_Fv\\ \alpha^i_F&b^i_F\tilde v
\end{array}\right),
\eeq
where $i=1$ or $2$ or $3$.
Above matrices can be diagonalized to
\beq
M_f^{Di}=\left(\begin{array}{cc}m^i&0\\ 0&M^i\end{array}\right)\simeq
\left(\begin{array}{cc}{{h^i_Fv\alpha^i_F}/{b^i_F\tilde v}}&0\\0&b^i_F
\tilde v\end{array}\right)
,\eeq
where $M^i$ represents heavy fermion masses.
Since the mass parameter $\alpha_F$ can be of order $\tilde v$, the
light fermion masses are of order $v\times$ (coupling constants).
We note that the large ratio of Yukawa couplings $f_t/f_u$ in the
standard model can be distributed to two classes of
couplings $h_F^i$ and $\alpha_F^i$, and hence lessening the
fermion mass hierarchy problem.
Now, we can represent fermions as mass eigenstates $f'$ and $ F'$
\bea
f_L&\simeq& f'_L+ \frac{h^*_Fv}{b_F\tilde v}
F'_L,\ F_L\simeq F'_L-\frac{h_Fv}{b_F\tilde v}f'_L
\nonumber\\
f_R&\simeq& f'_R+ \frac{\alpha^*_F}{b_F\tilde v}F'_R,\ F_R
\simeq F'_R-\frac{\alpha_F}{b_F\tilde v}f'_R
.\eea
So, all light fermions $f'$ acquire their masses through the seesaw mechanism.
For the neutrinos, the mass matrix can
be written in $(\nu_L, N_R)\otimes(\nu_L, N_R)$
basis,
\beq
M_N=\left(\begin{array}{cc}0&h_Nv\\ h^*_Nv&b_N\tilde v\end{array}\right),
\eeq
Diagonalizing its mass matrix,
a light neutrino acquires its mass
$m_\nu\simeq |h_Nv|^2/M_N$ which is very small.
For light neutrinos, there do not exist the $\alpha$ couplings
present in  $e,u$
and $d$ fermions; thus neutrinos  do not have mass at order $v$.

Axion effective interaction $a/(32\pi^2
\tilde v) F\tilde{F}$ term comes from the heavy singlet
quark $Q$, which has no light partner.
The PQ current of the lagrangian (2) is
\bea
J_\mu^{PQ}&=&\tilde v\partial_\mu a -\frac{1}{2}\left(\bar{Q}\gamma_\mu
\gamma_5 Q\right.
+\sum_{i=1}^3 [\bar{U}^i\gamma_\mu\gamma_5 U^i
+\bar{D}^i\gamma_\mu\gamma_5 D^i+\bar{E}^i\gamma_\mu\gamma_5 E^i
\nonumber\\
&&\left.-\bar{u}^i\gamma_\mu\gamma_5 u^i-\bar{d}^i\gamma_\mu\gamma_5 d^i
-\bar{e}^i\gamma_\mu\gamma_5 e^i]\right)
.\eea
The divergence of $J_\mu^{PQ}$ has the anomaly term only
\beq
\partial^\mu J_\mu^{PQ} = -\frac{a}{32\pi^2
\tilde v} F^a_{\mu\nu}\tilde{F}^{a\mu\nu}
.\eeq
If it were not for the anomaly term, $a$ would be massless.
To remove the anomaly
term, a proper axion current is defined as,
\beq
J_\mu^a=J_\mu^{PQ}+\frac{1}{2(1+Z)}(\bar{u}\gamma_\mu\gamma_5 u+Z
\bar{d}\gamma_\mu\gamma_5 d),
\eeq
where $Z=m_u/m_d$ and we neglect heavier quarks.
In the diagonalized basis, the axion does not mix with $\pi^0$
and $J^a_\mu$ has a divergence
\beq
\partial^\mu J_\mu^a=\frac{im_u}{1+Z}(\bar{u}\gamma_5 u+\bar{d}\gamma_5 d).
\eeq
Thus the axion mass is estimated as
\beq
m_a =\frac{f_\pi m_{\pi^0}}{\tilde v}\frac{\sqrt{Z}}{1+Z}
,\eeq
which is the same as that of the original heavy quark axion model.
However, the couplings between axion and matter fields are
rather different.
For the original model, the heavy quark axion couples to the light quarks
with the same strength \cite{kim}.
But, this new heavy quark axion has different couplings between $u$ and
$d$ quarks,
\bea
<\beta, a|{\cal L}|\alpha>&=& \frac{-i}{\tilde v}<\beta|[Q^a_5, {\cal L}]
|\alpha>\nonumber\\
&\simeq&
im_u \frac{2+Z}{1+Z}<\beta|\bar{u}\gamma_5 u
|\alpha>+im_d \frac{1+2Z}{1+Z}<\beta|\bar{d}\gamma_5 d|\alpha>.
\eea
For the leptonic sector, the original heavy quark axion
has no tree level axion
lepton coupling, and arises at one loop order. In this new model, below
symmetry breaking scale,
we have tree level lepton-axion coupling,
\beq
{\cal L}_{lla}=
\frac{ia}{\sqrt{2}}\frac{m_e}{\tilde v}\bar{e}^i \gamma_5 e^i.
\eeq
\begin{table}[h]
\begin{tabular}{cccc}
\hline\hline
Couplings & DFSZ axion & Hadronic axion & New heavy quark axion\\ \hline
&&&\\
$c_{a\gamma\gamma}$ &$0.75$ &$q^2-1.92$ &$q^2-1.92$ \\
&&&\\
$g_{aee}$ &$1.4\times 10^{-11}X_d\left(\frac{3}{N_g}\right)\left(
\frac{m_a}{\rm eV}\right)$&$\sim 0$ &$5.8\times 10^{-11}
\left(\frac{m_a}{\rm eV}\right) $ \\
&&&\\
\hline\hline
\end{tabular}
\caption{Photon--axion, electron--axion couplings \protect\cite{kim}.
Here $Z=0.56$,
and electromagnetic charge of $Q$ is defined as $qe$.}
\end{table}

In this paper, we have explored the consequences of new heavy quark axion
model.
The main idea of this model is to combine the two different scales,
the axion decay constant and the right-handed neutrino mass in the
 seesaw model.  Our model is a
simple extension of the standard model by doubling the number of fermions
and with additional $U(1)_{PQ}$ symmetry.
It is consistent with the existing experimental
data, since below the electroweak scale it reproduces the standard
model except for the very light majorana neutrinos which solve the solar
neutrino puzzle and the invisible axion which solves the
strong CP problem.
In contrast to the original heavy quark axion, new heavy quark axion
has tree level axion lepton coupling term and different coupling strengths
between $u$ and $d$ quarks.
It will be interesting to see if our model can be extended to a GUT.
More careful experiments are required to distinguish the two Higgs doublet
axion model and the one Higgs doublet axion model
(the heavy quark axion model).
\vskip 1.5cm

\noindent{\bf Acknowledgments}\\

This work is supported in part by Korea Science and Engineering
Foundation through Center for Theoretical Physics,
Seoul National University, and KOSEF--DFG Collaboration Program.

\end{document}